\documentclass[12pt]{article}  
\pdfoutput=1
\usepackage{graphicx}  
\usepackage{dcolumn}   
\usepackage{bm}        
\usepackage{amssymb}   
\usepackage{color} 
\usepackage{scicite}

\newcommand{\highlitq}[1]{{}} 

\usepackage{scicite}

\newenvironment{sciabstract}{%
\begin{quote} \bf}
{\end{quote}}

\newcounter{lastnote}
\newenvironment{scilastnote}{%
\setcounter{lastnote}{\value{enumiv}}%
\addtocounter{lastnote}{+1}%
\begin{list}%
{\arabic{lastnote}.}
{\setlength{\leftmargin}{.22in}}
{\setlength{\labelsep}{.5em}}}
{\end{list}}

\title{Probing spin-charge separation in a Tomonaga-Luttinger liquid}

\author
{Y. Jompol,$^{1,\ast}$ C. J. B. Ford,$^{1}$ J. P. Griffiths,$^{1}$
\\
I. Farrer,$^{1}$ G. A. C. Jones,$^{1}$ D. Anderson,$^{1}$ D. A. Ritchie,$^{1}$
\\
T. W. Silk$^{2}$ and A. J. Schofield$^{2}$\\
\\
\normalsize{$^{1}$Cavendish Laboratory, University of Cambridge,} \\
\normalsize{J J Thomson Avenue, Cambridge CB3 0HE, UK}\\
\normalsize{$^{2}$School of Physics and Astronomy, University of Birmingham,} \\
\normalsize{Edgbaston, Birmingham B15 2TT, UK}\\
\\
\normalsize{$^\ast$Present address: Nanoscience Technology Center, University of Central Florida,}\\
\normalsize{12424 Research Parkway Suite 400, Orlando, FL 32826, USA. To whom}\\
\normalsize{correspondence should be addressed. E-mail: yodchay.jompol@cantab.net}
}

\date{}

\begin{document}

\maketitle
\begin{sciabstract}
In a one-dimensional (1D) system of interacting electrons, excitations of spin and charge travel at different speeds, according to the theory of a Tomonaga-Luttinger Liquid (TLL) at low energies. However, the clear observation of this spin-charge separation is an ongoing challenge experimentally. We have fabricated an electrostatically-gated 1D system in which we observe spin-charge separation and also the predicted power-law suppression of tunnelling into the 1D system. The spin-charge separation persists even beyond the low-energy regime where the TLL approximation should hold. TLL effects should therefore also be important in similar, but shorter, electrostatically gated wires, where interaction effects are being studied extensively worldwide.
\end{sciabstract}

The effects of interactions are almost impossible to calculate in a general many-particle system, though they cannot be ignored. However, for a one-dimensional (1D) system, Luttinger, building on an approximation scheme of Tomonaga, constructed a soluble 1D model with infinite linear dispersion and a restricted set of interactions. The solution has the remarkable property that the excitations of spin and charge behave independently and move with different speeds.
It has been argued\cite{Haldane} that all 1D metals are adiabatically continuous with the Tomonaga-Luttinger model at low energies, and hence spin-charge separation should be observable in real systems.
Determining the extent of its applicability would provide a major test of more general methods of modelling interaction effects, with relevance to quantum devices and the theory of high-temperature superconductivity. Recent work \cite{Imambekov09} presents a more general theory of 1D systems with a nonlinear dispersion, but the effects of spin are not yet included.

Some properties of the TLL, such as power-law behavior, have been observed and studied in a variety of systems such as carbon nanotubes\cite{Bockrath} and edge states in the fractional quantum Hall regime \cite{AMChang}, but these experiments have not directly resolved the dispersion of the excitations in a TLL.
Only a few experiments have attempted to detect the spin-charge separation directly, e.g.\ by photoemission spectroscopy\cite{Kim,Kim06,Segovia} and tunnelling spectroscopy between a pair of closely spaced cleaved-edge-overgrown quantum wires \cite{Tserkovnyak,Auslaender02,Auslaender05}. The latter 1D-1D tunnelling results are striking and provide some evidence of dispersing spinons and holons -- the excitations of a TLL. However, in these experiments TLLs act as both probe and subject, so an independent study---in a different geometry---of the excitation spectrum is vital to be sure of the interpretation.

We use a two-dimensional electron gas (2DEG) as the (well-understood) probe layer. Use of an array of highly regular wires averages out impurity, length-resonance and charging effects, and measurements of power-law behaviour and spin-charge separation can be made with just the lowest 1D subband, without electrons becoming localised. Interpretation of TLL results is thus much easier, and results obtained in the nonlinear regime, where the theory is much less well developed, can be directly interpreted as a modification of the 1D elementary excitations. Our 1D wires are formed using split gates, where the confinement is much weaker than in the overgrown wires, making the results relevant to a broad range of other devices.
The tunnelling current $I$ between the 1D wires and an adjacent low-disorder layer containing a two-dimensional electron gas (2DEG) depends on the overlap between the spectral functions of the two systems.  This overlap is varied by using an in-plane magnetic field $B$ perpendicular to the wires to offset the two spectral functions in $k$-space by $\Delta k=eBd/\hbar$ along the wires, where $d$ is the centre-to-centre tunnelling distance between the two systems\cite{Kardynal}.
By applying a positive bias $V_{\rm dc}$ to the 2DEG relative to the wires, electrons tunnel into excited states of the 2DEG, from matching states below the Fermi energy in the 1D wires, allowing investigation of the energy dependence. Thus the 2D system acts as a spectrometer, and the differential conductance $G$ displays resonant-tunnelling peaks corresponding to overlapping points in the offset dispersion relations, where energy and momentum are conserved.

\begin{figure}[h] 
\begin{center}\leavevmode
\includegraphics[width=0.7\linewidth]{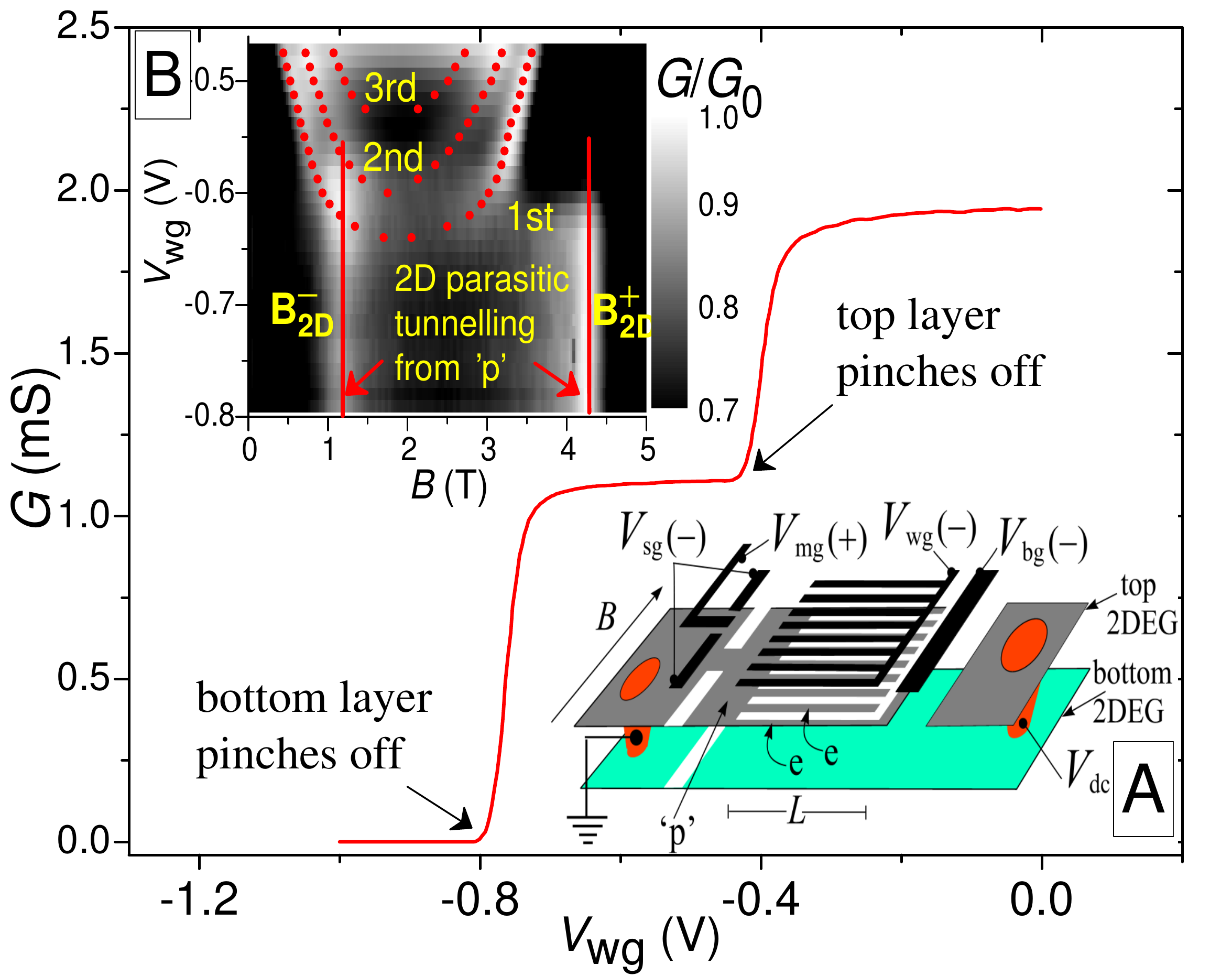}
\caption{The conductance, $G$, as a function of the wire-gate voltage $V_{\rm wg}$ (device A) with no other gates defined. It shows that the strip joining all the wire gates together (see Inset {A}) is depleted at $-0.4$~V, blocking conduction along the top layer; electrons then have to tunnel to the bottom layer, which is only depleted at $-0.8$~V. Inset \textbf{A}, device layout showing gate positions in black; the signs beside the voltage symbols indicate the polarity of the various gate voltages. Current is injected into both layers at the left Ohmic contact, then three gates pinch off the lower layer so that electrons flow into the wires in the upper layer, tunnel to the lower 2DEG and then to the right contact. Inset \textbf{B}, grey-scale plot of the tunnelling conductance showing resonant peaks (bright) as a function of transverse magnetic field measured at a lattice temperature of $\sim 40$~mK, with each trace normalised by the maximum height $G_0$. The dots indicate the features corresponding to the first, second and third 1D subbands, as labelled, measured from the raw data. The 2D parasitic tunnelling from region `p' does not change with $V_{\rm wg}$---its peaks stay at $B^-_{\rm 2D}=1.27$~T and $B^+_{\rm 2D}=4.31$~T, as indicated with vertical solid lines.} \label{figsubbands}\end{center}\end{figure}

\begin{figure}[h] 
\begin{center}\leavevmode
\includegraphics[width=0.7\linewidth]{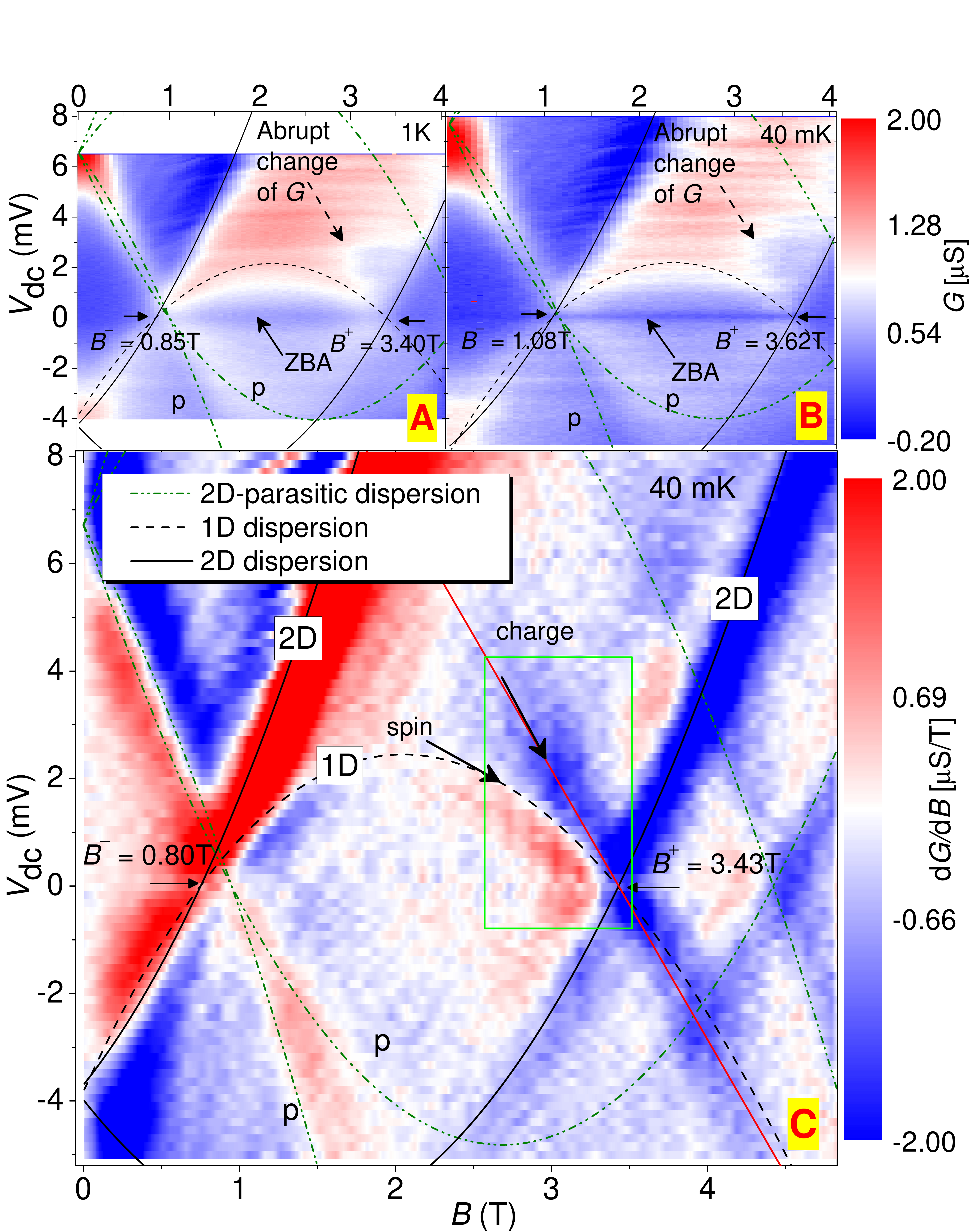}
\caption{\textbf{A},\textbf{B}, Colour-scale plots of $G$ \emph{vs} $V_{\rm dc}$ and $B$ at lattice temperatures of 1~K and 40~mK respectively (device A, for $V_{\rm wg}=-0.62$~V). Black lines (solid and dashed) indicate the locations of singularities predicted by the non-interacting model for tunnelling between the wires and the 2DEG, while the green dash-dotted lines indicate the locations of the singularities associated with the parasitic 2D-2D tunnelling (see main text). There is an additional abrupt decrease in $G$ along the line indicated. In addition, $G$ is suppressed at zero bias, labelled ZBA; this is another sign of interactions. \textbf{C}, ${\rm d}G/{\rm d}B$ (device A, for $V_{\rm wg}=-0.60$~V.
The non-interacting parabolae are shown as in \textbf{A} and labelled 1D or 2D to indicate which dispersion is being probed. The straight red line indicates the locus of the abrupt change in $G$ indicated in \textbf{A} and \textbf{B}, and is a factor of $\sim 1.4$ steeper than the 1D parabola at $V_{\rm dc}=0$. Importantly, this feature clearly moves away from the 1D parabola.  We identify it with the TLL charge excitation (holon), while the 1D parabola tracks the spin excitation (spinon).
} \label{GvsBV}\label{dGdB}\end{center}\end{figure}

Our devices contain an array of about 350 extremely regular quantum wires in the upper layer of a GaAs-AlGaAs double-quantum-well (DQW) structure (see Fig.\ \ref{figsubbands}{A}). The two layers are separately contacted using a surface-gate depletion scheme\cite{Nield,JompolEP2DS,SOM}. The wires, of length $L=17.5~\mu$m and lithographic width 0.17~$\mu$m (device A) and 0.18~$\mu$m (device B), were squeezed by a negative gate voltage $V_{\rm wg}$. There is an additional small ungated region `p' of width $0.9~\mu$m, which provides a current path to the entrances to the 1D wires. The tunnelling conductance $G = {\rm d}I/{\rm d}V_{\rm dc}$ between the 1D wires and the 2D layer was measured as a function of source-drain bias $V_{\rm dc}$ in an in-plane magnetic field $B$ perpendicular to the wires at lattice temperatures $T$ down to $\sim 40$~mK. 

\begin{figure*} 
\begin{center}\leavevmode
\includegraphics[width=\textwidth]{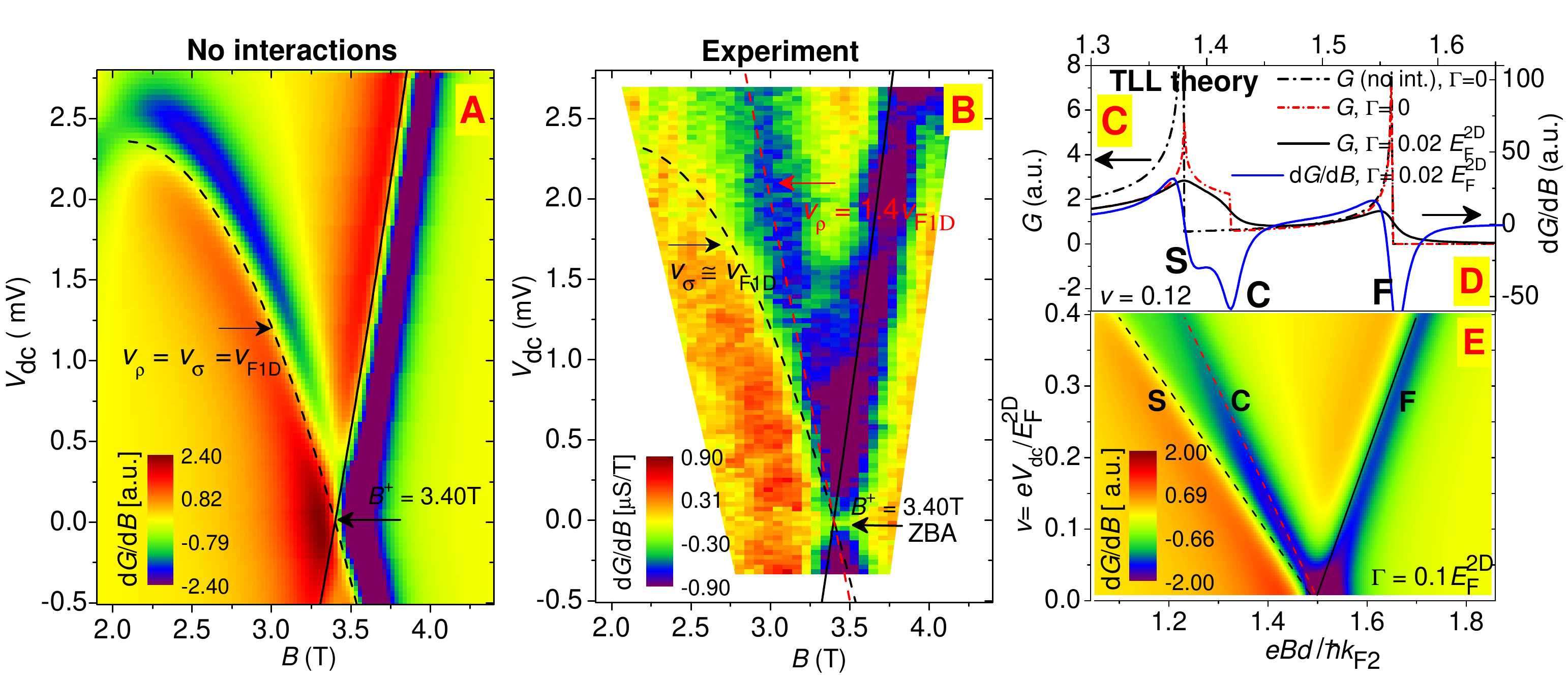}
\caption{Comparison of ${\rm d}G/{\rm d}B$ for experiment and theory. \textbf{A}, For non-interacting electrons, all features track the non-interacting parabolae (disorder broadening $\Gamma=0.6$~meV); \textbf{B}, $\mathrm d G/\mathrm dB$ measured at high resolution while sweeping $B$, for Device B. The red line marks a feature which does not track the non-interacting parabolae, and is absent in \textbf{A}.
Calculation of $G$ (\textbf{C}) and ${\rm d}G/{\rm d}B$ (\textbf{D}) for non-interacting electrons and a TLL.
The dimensionless bias $v=eV_{\rm dc}/E_{\rm F}^{\rm 2D}=0.12$; $\Gamma$ is indicated for each curve. For the TLL, the spinon velocity $v_{\sigma} = v_{\rm F1D}$ and the chosen holon velocity $v_{\rho} = 1.5v_{\rm F1D}$. Spin and charge excitations are labelled S and C respectively. F labels the non-interacting 2D dispersion curve. \textbf{E} ${\rm d}G/{\rm d}B$ as a function of $B$ and $v$, showing the same charge feature (C) as in the experiment (\textbf{B}).}
\label{dGdBexpttheory}\end{center}\end{figure*}

Firstly, the effect of the wire-gate voltage $V_{\rm wg}$ was characterised (see Fig.\ \ref{figsubbands}). Initially, at small $V_{\rm wg}$, the upper 2DEG screens the lower 2DEG. The first drop in $G$ occurs at $-0.40~$V when the top well depopulates, leaving narrow 1D wires between the gates. Then the bottom well is gradually depleted until it pinches off at $-0.80~$V. $B$ was then swept at each $V_{\rm wg}$ (with $V_{\rm dc} = -0.3$~mV, to avoid the zero-bias anomaly described later), revealing a series of peaks (Fig.\ \ref{figsubbands}{B}), one pair of peaks at $B_i^\pm$ for the $i$-th 1D subband ($i \ge 1$). For this device (A, $0.17~\mu$m wire width), six peaks are observed at $V_{\rm wg} = -0.49$~V, corresponding to three occupied 1D subbands\cite{JompolEP2DS}. With more negative $V_{\rm wg}$, the 1D subband spacing increases and the 1D density decreases. Hence the number of subbands can be reduced to just one. At $V_{\rm wg} {\sim} -0.65$~V the wires become insulating. The widths of the wire gates and of the long narrow 2D `p' region were chosen carefully such that even with just a single 1D subband in the top 2DEG there was minimal modulation of the lower 2DEG and current could still reach the wires. However, the `p' region inevitably contributes a 2D-2D parasitic tunnelling current that appears in the measurements, but this is small and independent of the tunnel current from the wires, even after wire pinch-off, so it can be measured and allowed for.

We now choose $V_{\rm wg}= -0.62$~V, well into the region where there is just one 1D subband. Fig.\ \ref{GvsBV}{A} and {B} show $G$ as a function of the dc inter-layer bias ($V_{\rm dc}$) and $B$ as colour-scale plots, at high (1~K) and low ($\sim 40$~mK) lattice temperatures respectively. In the absence of interactions, there should be peaks in $G$ that track the 1D and 2D parabolic dispersion relations; these are indicated with dashed and solid black curves, respectively. The parasitic 2D-2D tunnelling in the ungated `p' region is also shown, as green dash-dotted lines labelled p. The crossing points along $V_{\rm dc} = 0$ occur at $B^{-} = 1.08$~T and $B^{+} = 3.62$~T (see Fig.~\ref{dGdB}{B}). The 1D Fermi wave vector is $k_{\rm F1D}=ed(B^{+}-B^{-})/2\hbar$, giving the approximate electron density in the wires $n_{\rm 1D} \simeq 40 \mu$m$^{-1}$ (from $k_{\rm F1D}=\pi n_{\rm 1D}/2$).
The upturned (dashed) parabola maps the energy of the 1D wires as a function of wave vector $k$. 

The curves drawn in Fig.\ \ref{GvsBV} are those expected for single-particle tunnelling. However, there is an additional region of high conductance to the right of $B^+$. It drops off along a straight line moving diagonally up and to the left from $B^+$ (indicated with an arrow). To show this more clearly, the data are differentiated with respect to $B$ and plotted in Fig.\ \ref{dGdB}{C}. We have taken detailed data in this region, sweeping $B$, in another device (B) in the same single-subband regime (see Fig.\ \ref{dGdBexpttheory}B). We find a region of large negative ${\rm d}G/{\rm d}B$ along a straight line to the right of the high-conductance region; this appears dark blue and is indicated by the red dashed line.  We compare this with the theoretical prediction for non-interacting electrons, shown in Fig.\ \ref{dGdBexpttheory}A, where a similar dark blue feature also occurs. Importantly, however, it tracks the 1D parabola along its length.  In contrast, the feature in Fig.\ \ref{dGdBexpttheory}{B} disperses away from the 1D parabola.  As features in the conductance reflect singular features in the spectral function we can conclude that the 1D parabola and the red dashed line track the dispersion of two independent excitations, which in the TLL framework correspond to the spinon and holon, for spin and charge excitations, respectively.

To confirm this interpretation, we have calculated the tunnelling spectra for a TLL coupled to a 2D system of electrons.  The framework for these calculations already exists in the literature \cite{Altland,Grigera}, so we only describe the relevant details here.  To compute the tunnelling current, we require the spectral function of a TLL, which in general depends on four parameters: the spinon and holon velocities $v_{\sigma}$ and $v_{\rho}$, respectively, plus two exponents $\gamma_{\sigma}$ and $\gamma_{\rho}$. We assume repulsive, spin-rotation-invariant interactions, which implies $\gamma_{\sigma}=0$, and $v_\rho > v_\sigma$; in the absence of back-scattering $v_\sigma \approx v_{\rm F1D}$, the 1D Fermi velocity.  For simplicity, we also neglect interbranch scattering processes, which sets $\gamma_\rho=0$. The spectral function of the 2D system is taken to be a Lorentzian of width $\Gamma$, where $\Gamma$ is the disorder-scattering rate in the 2D system\cite{NTurner}.

This minimal model is convenient for analytical calculations, and is expected to reproduce the main features of the tunnelling spectra associated with spin-charge separation.  It will not, however, capture the zero-bias anomaly, which is absent for $\gamma_{\rho}=0$.  Finite $\gamma_{\rho}$ is also expected \cite{Grigera} to lead to a weakening of the singular features in the spectra, making our calculation the `optimum' case for observing spin-charge separation.  Since the TLL model relies on the linearisation of the spectrum at low energies, the model's results are only formally applicable in the low-bias part of the spectrum, where this linear approximation holds.

Figure \ref{dGdBexpttheory}{C} shows the calculated conductance in the vicinity of the $B^+$ point as a function of scaled magnetic field at fixed voltage, both for non-interacting electrons, a clean TLL, and a TLL with a little disorder broadening. The singular peak at high magnetic field is independent of the form of the excitations, and hence is common to both systems.  The peak at lower magnetic field for the non-interacting system splits, in the TLL case, into a weakened singular peak (which for $v_{\sigma} = v_{\rm F1D}$ occurs at the same magnetic field as the non-interacting peak), plus a finite discontinuity away from the peak.  This sudden drop in the conductance away from the peak is precisely what is observed experimentally in figures \ref{dGdB}A and B.  Fig.\ \ref{dGdBexpttheory}D shows $\mathrm{d}G/\mathrm{d}B$ with a little disorder. The spinon is identified as a maximum/minimum pair, which disperse together in the $B$-$V_{\rm dc}$ plane, as shown in Fig.\ \ref{dGdBexpttheory}E, for more realistic disorder broadening.  The holon is identified as a single minimum, which disperses away from the spinon.  We note that an extra feature is also predicted to occur at the $B^{-}$ point. However, since all three features from this point remain in proximity to each other, it is difficult to resolve them individually.

The experimental results in Fig.\ \ref{dGdBexpttheory}{B} are consistent with the predictions in Fig.\ \ref{dGdBexpttheory}{E}, at least in the low-bias regime where the linear approximation is reasonable.  We have observed this extra feature in the conductance in three devices, at temperatures between $\sim 60$~mK and 1.8~K, and on several thermal cycles. We thus conclude that we are observing spin-charge separation, the hallmark of a TLL.

\begin{figure}[h] 
\begin{center}\leavevmode
\includegraphics[width=0.7\linewidth]{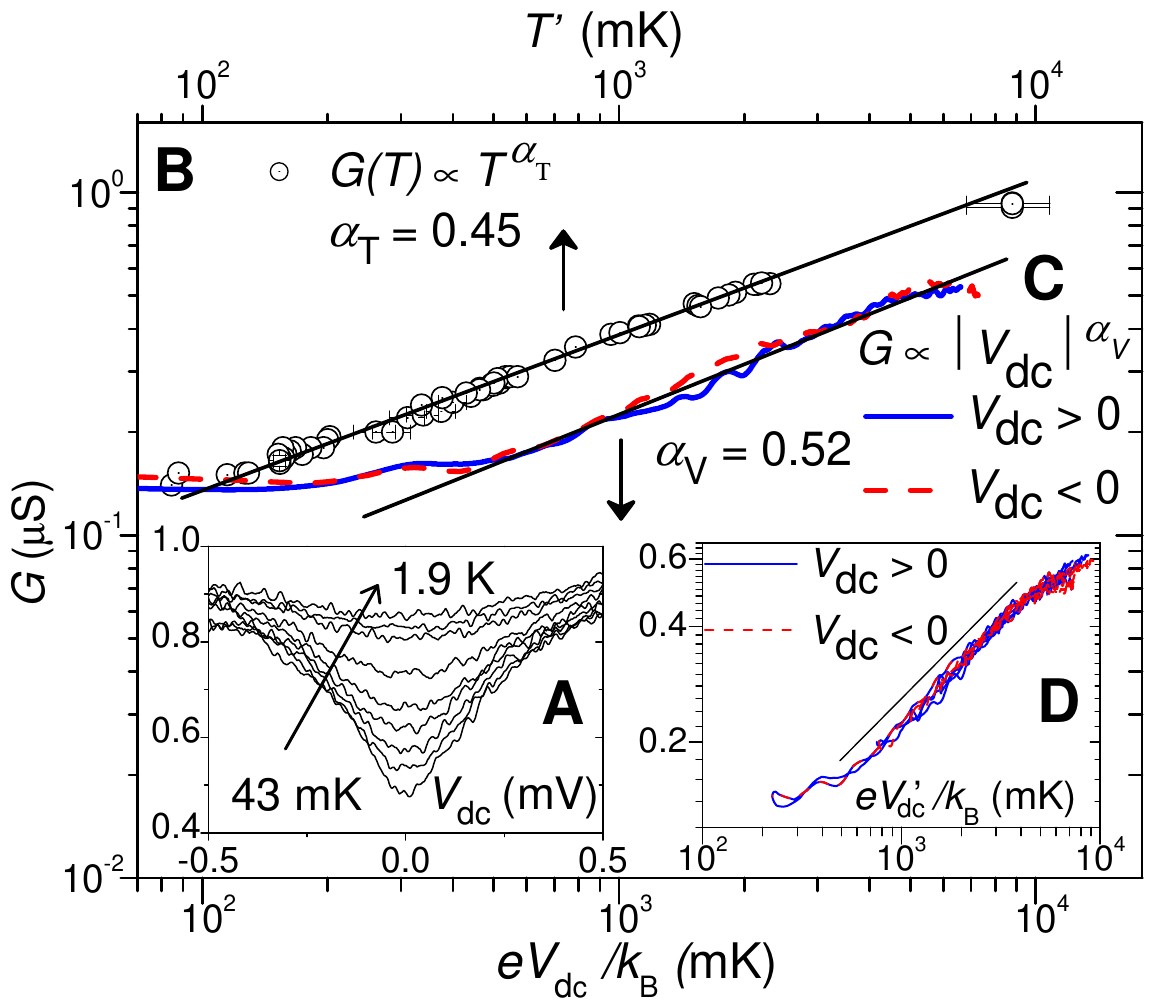}
\caption{The dependence of the conductance $G$ on bias $V_{\rm dc}$ and temperature $T$ (device B). \textbf{A}, The ZBA at various $T$ up to 1.9~K at $B=2.33$~T.
\textbf{B}, Minima of the ZBA ($G(V_{\rm dc}=0)$) \emph{vs} $T$. $G$ varies as $T^{\alpha_{\rm T}}$ where $\alpha_{\rm T} \approx 0.45$ over nearly two orders of magnitude. \textbf{C}, $G(|V'_{\rm dc}|)$ on a log-log plot (with $V_{\rm dc}$ scaled to temperature using $eV_{\rm dc}=k_{\rm B}T$), on scales matching those in \textbf{B} for $T$, showing that it varies as $V_{\rm dc}^{\alpha_{\rm V}}$ where $\alpha_{\rm V} \approx 0.52$ over more than one order of magnitude. \textbf{D}, $G(|V_{\rm dc}|)$ \emph{vs} $V'_{\rm dc}=\sqrt{|V_{\rm dc}|^2+(3k_{\rm B}T/e)^2+V_{\rm ac}^2}$, the simplest way of adding the forms of energy smearing together as noise in quadrature. The value $3k_{\rm B}T$ is chosen to superpose the lines in \textbf{B} and \textbf{C} and also all the curves in \textbf{D}, where the lowest temperature has been adjusted from 43~mK to 70~mK to avoid saturation. Using the more complex scaling relation of equation 5 of ref. \citen{Bockrath} also gives a universal curve if $\alpha \approx 0.51$. 
} \label{ZBA}\end{center}\end{figure}

An additional feature that cannot be explained in the
non-interacting model is the `zero-bias anomaly' (ZBA), the strong
suppression of conductance along $V_{\rm dc}=0$, visible as a dark-blue
line in Fig.\ \ref{GvsBV} {A} and {B}. This is likely to be related to the
energy cost for an electron to tunnel into or out of a 1D wire, as
it disturbs the line of electrons on either side of it. It has
previously been observed in 1D-1D tunnelling\cite{Tserkovnyak} and,
for a TLL, $G$ is expected to have identical power-law dependences
on $V_{\rm dc}$ and temperature $T$. 

Fig.\ \ref{ZBA}{A} shows the ZBA in the tunnelling conductance $G$ as a function of $V_{\rm dc}$ at various $T$, at a field midway between $B^-$ and $B^+$. Fig.\ \ref{ZBA}{B} and {C} show $G(V_{\rm dc}=0,T)$ and $G(|V_{\rm dc}|,T<70~{\rm mK})$, respectively, on log-log plots. Both clearly vary as power laws, as labelled, over considerably more than one order of magnitude. The corresponding power-law exponents $\alpha_{\rm T} \approx 0.45\pm0.04$ and $\alpha_{\rm V} \approx 0.52\pm0.04$ are very similar, as expected. To illustrate that temperature and bias play a similar role in smearing the energy, $G(|V_{\rm dc}|)$ is plotted in Fig.\ \ref{ZBA}{D} as a function of $V'_{\rm dc}=\sqrt{(|V_{\rm dc}|^2+(3k_{\rm B}T/e)^2+V_{\rm ac}^2)}$, the simplest way of adding forms of energy smearing together as noise in quadrature. ($3k_{\rm B}T$ is an estimate of the thermal energy spread, and the factor of $3$ is chosen to match the offset between the curves in Fig.\ \ref{ZBA}{B} and {C}.)

The TLL is characterized by the `anomalous' exponent $\gamma_\rho$ and the spinon and holon velocities. $\gamma_\rho$ can be expressed in terms of an interaction parameter $g$ ($\gamma_{\rho}=(g + 1/g -2)/8$) which indicates the strength of the interactions ($g=1$ for non-interacting particles). Accurate {\it a priori} calculation of $g$ cannot be done in general. However an estimate based on the 1D densities used suggests $g\sim 0.7$ (supporting online text). We now determine $g$ from the various experimental results.

In TLL theory, the spin velocity $v_\sigma$ is approximately equal to the Fermi velocity $v_{\rm F1D}$ for weak backscattering ($v_\sigma < v_{\rm F1D}$ for stronger interactions) \cite{VoitPRB,Creffield}. Thus the spin mode should follow the low-bias part of the non-interacting 1D parabola (or even lie below it). The charge mode should propagate faster than the spin mode for repulsive interactions. Thus we label the curve that follows the 1D parabola `spin' and the extra line `charge'. For device B, we deduce the velocities close to zero bias to be $v_\sigma \approx v_{\rm F1D} = \hbar k_{\rm F}/m_{\rm 1D}=1.13\times 10^5$~ms$^{-1}$ and $v_\rho \approx 1.4 v_{\rm F1D}$, respectively. In a Galilean invariant system $g=v_{\rm F1D}/v_\rho$~\cite{Creffield}, so this gives $g \approx 0.7$, which is consistent with the estimate given above. For Device A, $g$ is very similar to that of Device B, as is the 1D electron density. Another device, with a thinner (7 nm) barrier, yields $g \approx 0.65$.

Within TLL theory, the power-law exponents extracted from the zero-bias anomaly can also be directly related to $g$. The form of the exponent depends on whether or not the excitations are significantly affected by the ends of the wire. This is determined by comparing the energy of the tunnelling electron to $\Delta E = 2\hbar v_{\rm F1D}/gL$, which is related to the inverse timescale for the holon to travel to the end of a wire of length $L$\cite{KaneFisher,Tserkovnyak}. For $k_{\rm B}T, eV_{\rm dc} \gg \Delta E$, the ends are unimportant and the process is called `bulk' tunnelling, with an exponent $\alpha_{\rm bulk} = (g+1/g-2)/4$. Taking $L = 17\mu$m and $g=0.7$ gives $\Delta E \approx 150$mK for our device, and therefore almost all of our ZBA data shown in Fig. \ref{ZBA} are expected to be in the bulk-tunnelling regime. The measured values of $\alpha_{T} = 0.45$ and $\alpha_{V}=0.52$ (device B) give $g \approx 0.28$ and $0.26$, respectively.

The values of $g$ found from the ZBA exponents are significantly smaller than that extracted from the holon branch in Fig. \ref{dGdBexpttheory}. We briefly offer possible explanations for this (with more detail in the S.O.M.). One possibility is that impurities and imperfections may make the effective length of each wire shorter than the lithographic length. It is well known that an impurity in a TLL will effectively cut it into two at low energies. Although at finite energies, electrons are able to tunnel through such constrictions, one might expect that the constrictions will modify the form of the ZBA exponent in a similar way to `ends'. If we use the formula for ZBA exponent in the end-tunnelling regime \cite{SOM}, the extracted values of $g$ (0.53 from $\alpha_T$ and 0.49 from $\alpha_{V}$) are more comparable to that extracted from the holon branch.

Alternatively it may not be valid to extrapolate the higher-energy properties of the branches to low energies where the ZBA is measured. The interactions between the wires and/or between the wires and the 2DEG may become significant, and may alter the form of the ZBA exponent\cite{CarpentierPRB}. The effective dielectric constant of the material may also be energy dependent, changing the strength of the interactions at low energies.

TLL theory is based on the assumption of a linear dispersion relation about the Fermi energy. We go beyond that regime at high DC bias. There is very little work on interactions in 1D wires at high energy where the TLL approximation breaks down. Haldane\cite{Haldane} argued that all 1D metals are adiabatically continuous with the TLL, i.e.\ perturbations such as the band curvature are only expected to lead to a renormalisation of the TLL parameters, so spin-charge separation should persist; this is backed up by recent renormalisation-group calculations \cite{Benthien}. Recent theoretical work on spinless fermions for a curved band has shown an intriguing interplay of fermi-liquid and TLL behaviour\cite{Khodas,Imambekov09}. Numerical calculations have been performed in this regime\cite{Zacher} using quantum Monte Carlo methods, with results that resemble our experimental results---a parabolic spin branch and a fairly straight charge branch.

\bibliography{1171769}		

\bibliographystyle{Science}	

\begin{scilastnote}
\item We acknowledge the UK EPSRC for funding; YJ was supported by a Scholarship from the Thai Ministry of Science. We thank F.\ Sfigakis for experimental assistance and B.E.\ Kardynal and C.H.W.\ Barnes for useful discussions.
\end{scilastnote}

\textbf{Supporting Online Material}

www.sciencemag.org

Materials and Methods

SOM Text

\end{document}